\documentclass[11pt]{article}
\usepackage{epsfig}
\def\Tr{{\rm Tr}}
\def\ln{{\rm ln}}

\def\d {\partial}

\def\pr{Phys. Rev. }
\def\np{Nucl. Phys. }
\def\pl{Phys. Lett. }
\def\zp {Z. Phys. }

\newcommand{\la}[1]{\label{#1}}

\newcommand{\ed}{\end{document}}
\newcommand{\bq}{\begin{equation}}
\newcommand{\eq}{\end{equation}}
\newcommand{\ba}{\begin{eqnarray}}
\newcommand{\ea}{\end{eqnarray}}
\newcommand{\baz}{\begin{eqnarray*}}
\newcommand{\eaz}{\end{eqnarray*}}
\newcommand{\bb}{\begin{thebibliography}}
\newcommand{\eb}{\end{thebibliography}}
\newcommand{\ct}[1]{${\cite{#1}}$}

\newcommand{\bi}[1]{\bibitem{#1}}

\newcommand{\lsim}{\stackrel{<}{\sim}}
\newcommand{\ben}{\begin{enumerate}}
\newcommand{\een}{\end{enumerate}}
\begin{document}
\begin{titlepage}
%\centerline{\large Preliminary}

\hfill {FTUV 98-74; IFIC 98-75}

\hfill {KIAS-P98030}

\hfill {SNUTP/98-124}

\hfill {\today: hep-ph/9810539}

\begin{center}
\ \\
{\Large \bf The Proton Spin in the Chiral Bag Model:} \\ {\large
\bf Casimir Contributions and Cheshire Cat Principle}
\ \\
\vspace{.3cm}
{Hee-Jung Lee and Dong-Pil Min }

{\it Department of Physics and Center for Theoretical Physics}\\
{\it Seoul National University, Seoul 151-742, Korea}\\ ({\small
E-mail: hjlee@fire.snu.ac.kr; dpmin@phya.snu.ac.kr})

%                 and
\vskip 0.2cm
{Byung-Yoon Park}

{\it Department of Physics, Chungnam National University, Daejon,
Korea}\\ ({\small E-mail: bypark@chaosphys.chungnam.ac.kr})

%                 and
\vskip 0.2cm
{Mannque Rho}

{\it Service de Physique Th\'eorique, CE Saclay}\\ {\it 91191
Gif-sur-Yvette, France}\\ {\&}\\{\it School of Physics, Korea
Institute for Advanced Study, Seoul 130-012, Korea}\\ ({\small
E-mail: rho@spht.saclay.cea.fr})

%                 and

\vskip 0.2cm
{Vicente Vento }

{\it Departament de Fisica Te\`orica and Institut de F\'{\i}sica
Corpuscular}\\ {\it Universitat de Val\`encia and Consejo Superior
de Investigaciones Cient\'{\i}ficas}\\ {\it E-46100 Burjassot
(Val\`encia), Spain}\\ ({\small E-mail: vicente.vento@uv.es})

\end{center}
\vskip 0.2cm
\centerline{\bf Abstract}

The flavor singlet axial charge has been a source of study in the
last years due to its relation to the so called {\it Proton Spin
Problem}. The relevant flavor singlet axial current is anomalous,
i.e., its divergence contains a piece which is the celebrated
$U_A(1)$ anomaly. This anomaly is intimately associated with the
$\eta^\prime$ meson, which gets its mass from it. When the gauge
degrees of freedom of QCD are confined within a volume as is
presently understood, the $U_A(1)$ anomaly is known to induce color
anomaly leading to ``leakage" of the color out of the confined
volume (or bag). For consistency of the theory, this anomaly should
be canceled by a boundary term. This ``color boundary term"
inherits part or most of the dynamics of the volume (i.e., QCD). In
this paper, we exploit this mapping of the volume to the surafce
via the color boundary condition to perform a complete analysis of
the flavor singlet axial charge in the chiral bag model using the
Cheshire Cat Principle. This enables us to obtain the hitherto
missing piece in the axial charge associated with the gluon Casimir
energies. The result is that the flavor singlet axial charge is
small independent of the confinement (bag) size ranging from the
skyrmion picture to the MIT bag picture, thereby confirming the
(albeit approximate) Cheshire Cat phenomenon.
.

\vskip 0.25cm
\leftline{Pacs: 12.39-x, 13.60.Hb, 14.65-q, 14.70Dj}
\leftline{Keywords:  quarks, gluons, mesons, anomaly, spin, proton}
%\vspace{0.5cm}

\end{titlepage}

\section{Introduction}

The possibility of formulating a physical theory by means of
equivalent field theories defined in terms of different field
variables, leads to a construction principle for
phenomenologically sensible and conceptually powerful models,
referred to as the Chesire Cat Principle (CCP)\ct{ccp,ccpph}. In
1+1 dimensions fermionic theories are bosonizable \ct{two} and the
CCP can be made exact and transparent. Furthermore in the
supersymmetric world, the powerful nonrenormalizable theorems
allow a web of exactly equivalent theories to be established. In
the real four-dimensional nonsupersymmetric world, bosonization
with a finite number of degrees of freedom is not exact. However
based on the unproven ``theorem" of Weinberg \ct{wei}, it seems
possible to argue that the CCP should hold also in four
dimensions, albeit approximately. In view of recent developments
on establishing the network of dualities in what is believed to be
a fundamental theory (i.e., string theory) where one might say
that the exact CCP holds, the notion of a precise CCP in the real
world is no longer so preposterous. The aim of the present
investigation is to show the full consistency of the CCP in the
hadronic world for the case of the {\it Proton Spin}, which was
not fully satisfactorily established in our previous efforts in
this direction \ct{pvrb,pv,rv}\footnote{Note that in these papers,
we have shown that the CCP holds for non-zero bag radii but it
failed when the bag radius shrank to a point, implying that in the
model studied, the pure skyrmion and the MIT bag did not have the
equivalent structure required by the CCP.}. We complete the
program in this paper.

Quantum Chromodynamics (QCD) is the theory of the hadronic
phenomena \ct{qcd}. At sufficiently low energies or long distances
and for a large number of colors $N_C$, it can be described
accurately by an effective field theory in terms of meson fields
\ct{mesons}. In this regime, the color fermionic description of
the theory is extremely complex due to confinement. However the
implementation of the CCP in a two phase scenario called the
Chiral Bag Model (CBM) has proven surprisingly powerful \ct{cbm}.

What is the CBM? Let space-time be divided in two regions by a
hypertube, that is, the evolving bag. In the interior of the tube,
the dynamics is defined in terms of the microscopic QCD degrees of
freedom, quarks and gluons. In the exterior, one assumes an
equivalent dynamics in terms of meson fields, i.e., one that
respects the symmetries of the original theory and the basic
postulates of quantum field theory \ct{wei}. The two descriptions
are matched by defining the appropriate boundary conditions which
implement the symmetries and confinement \ct{ccp,cbm}. What this
does effectively is to delegate all or part of the principal
elements of the dynamics taking place inside (QCD) the bag to the
boundary. We will see that this strategy works quite efficiently in
the problem at hand.

In this scenario the CCP states that the hadron physics should be
approximately independent of the spatial size of the confinement
region or the bag \ct{ccp}. This realization of the principle has
been tested in many instances in hadronic physics with fair success
\ct {ccpph}.

There is one case, however, where the realization of the CCP has
not been as successful as in the other cases, namely, the
calculation of the flavor singlet axial charge (FSAC) of the
nucleon. Indeed in the previous efforts \ct{pvrb,pv,rv}, the CCP
was realized only partially as it seemed to fail at certain points
such as for zero bag radius. It is the leitmotiv of this work to
remove this apparent failure.

The observable FSAC has become very relevant in the nucleon
structure in recent years, because it is associated with the so
called {\it Proton Spin Problem} \ct{spin}. The experimentally
observed small value for the FSAC implies a strong violation of the
so-called Ellis-Jaffe sum rule
\ct{ej} and therefore implies that the polarization of the proton
is not carried exclusively by the valence quarks. It is also very
interesting from the formal point of view, because the flavor
singlet axial current -- which is the origin of the observable --
is anomalous, and its anomaly is related in a non-trivial way with
the gluonic structure of the theory
\ct{anom}.

In the CBM, the scenario of how the CCP is realized -- which is the
central issue of our problem -- is very intricate. As stated, the
flavor singlet axial current is associated with the anomaly and
effectively with the $\eta^\prime$ meson. Thus, besides the pion
field of the conventional effective theories which accounts for
spontaneously broken chiral symmetry, the correct treatment of the
FSAC requires minimally the inclusion  of a field describing the
$\eta^\prime$ meson. We shall label it for brevity $\eta(x)$ since
no confusion will arise in what follows.

The intricacies of the {\it hedgehog} configuration and its
relevance to the fractionation of baryon charge and other
observables have been extensively discussed \ct{frac} and fairly
well understood \ct{pr,p}. They will be implemented in our
calculation without much details. Moreover the inclusion of the
$\eta^\prime$ meson carries subtleties of its own. The vacuum
fluctuations inside the bag, that induce the baryon number leakage
into the {\it skyrmion} \ct{frac}, also induce a color leakage if
a coupling to a pseudoscalar isoscalar field is allowed \ct{nrwz}.
This leakage would break color gauge invariance and confinement in
the model unless it is canceled. As suggested in \ct{nrwz}, this
color leakage can be prevented by introducing into the CBM
Lagrangian a counter term of the form \bq {\cal
L}_{CT}=i\frac{g^2}{32\pi^2}\oint_{\Sigma} d\beta K^\mu n_\mu
({\Tr}{ \ln} U^\dagger -{\Tr} {\ln} U)\label{lct} \eq where $N_F$
is the number of flavors (here taken to be =3), $\beta$ is a point
on a surface $\Sigma$, $n^\mu$ is the outward normal to the bag
surface, $U$ is the $U(N_F)$ matrix-valued field written as
$U=e^{i\pi/f} e^{i\eta/f}$ and $K^\mu$ the properly regularized
Chern-Simons current $K^\mu=\epsilon^{\mu\nu\alpha\beta} (G_\nu^a
G_{\alpha\beta}^a -\frac 23 f^{abc} g G_\nu^a G_\alpha^b
G_\beta^c)$ given in terms of the color gauge field $G^a_\mu$.
Note that (\ref{lct}) manifestly breaks color gauge invariance
(both large and small, the latter due to the bag), so the action
of the chiral bag model with this term is not gauge invariant at
the classical level but as shown in \cite{nrwz}, when quantum
fluctuations are calculated, there appears an induced anomaly term
on the surface which exactly cancels this term. Thus gauge
invariance is restored at the quantum level.

The equations of motion for the gluon and quark fields
inside and the $\eta^\prime$ field outside are the same as in
\ct{pvrb,pv}. However the boundary conditions on the surface with the
inclusion of Eq.(\ref{lct}) read \ct{rv}
\bq
\hat{n}\cdot \vec{E}^a=-\frac{N_F g^2}{8\pi^2 f} \hat{n}\cdot \vec{B}^a
\eta\label{E}
\eq
\bq
\hat{n}\times \vec{B}^a=\frac{N_F g^2}{8\pi^2 f} \hat{n}\times \vec{E}^a
\eta\label{B}
\eq
and
\bq
\frac 12 \hat{n}\cdot({\bar{\psi}}\vec{ \gamma}\gamma_5\psi)
=f \hat{n}\cdot\d \eta + \frac{N_F g^2}{16\pi^2 } \hat{n}\cdot K\label{bc}
\eq
where $\vec{E}^a$ and $\vec{B}^a$ are,
respectively, the color electric and color magnetic fields.
Here $\psi$ is the QCD quark field.

A complete treatment calls for a full Casimir calculation of the
gluon modes, which is highly subtle due to the p-wave structure of
the $\eta$-field. Such a calculation is in progress \ct{lmprv} and
will be reported in a later publication. Here we would like to
side-step this technically difficult procedure by first assuming
the CCP in evaluating the Casimir contribution with the color
boundary conditions (\ref{E}), (\ref{B}) and (\ref{bc}) taken into
account and check a posteriori that there is consistency between
the assumption and the result.

The next section will define our formulation, recall our old
results, and clarify the new contributions. Section 3 will focus on
the gluon Casimir contribution to the FSAC, the major contribution
of this presentation. Finally section 4 will contain the results,
conclusions and the future prospects of ongoing work.

\section{The Chiral Bag Formalism}

Our aim is to calculate the FSAC in the CBM scenario. In order to
do so we need a specific formulation of the model through its
equations of motion and boundary conditions. The equations of
motion have been shown repeatedly in our previous works
\ct{pvrb,pv,rv} and the color boundary conditions were recalled in
the introduction. We refer the reader to those references for a
detailed discussion on their structure, their resolution and the
implementation of gauge invariance and confinement. Our calculation
will be carried out in the static spherical cavity approximation,
that is, our bag will be a static sphere of radius $R$ dividing two
regions of space in which the theory is implemented by QCD for $r <
R$, and by an effective meson theory for $r > R $.

\subsection{The anomaly and proton spin}

The anomalous suppression of the first moment, $\Gamma^p_1$, of the
polarized proton structure function $g_1^p$ has been the focus of
intense theoretical and experimental activity for nearly a decade.
While it is now generally accepted that the key to understanding
this effect is the existence of the chiral $U(1)$ anomaly in the
flavor singlet axial current there are several explanations
reflecting different theoretical approaches to proton structure.

The starting point is the sum rule for the first moment, i.e.,
$$
\Gamma^P_1(Q^2) \equiv \int_0^1dx g_1^p(x,Q^2) =
\frac{1}{12}C_1^{NS}(\alpha_s(Q^2))\left(a^3
+ \frac{1}{3} a^8\right) $$
\bq
+\frac{1}{9}C_1^S(\alpha_S(Q^2))a^0(Q^2) .
\eq
Here $C_1(\alpha_s)$ are  first
moments of the Wilson coefficients of the the singlet ($S$) and nonsinglet
($NS$) axial currents and $\alpha_s$ the perturbatively running $QCD$ coupling
constant. Moreover
$a^3$, $a^8$ and $a^0(Q^2)$ are the form factors in the
forward proton matrix elements of the renormalized axial current,
i.e.,
$$
\langle p,s|A^3_\mu|p,s \rangle =s_\mu \frac{1}{2} a^3, \;\;\;\;\;
\langle p,s|A^8_\mu|p,s \rangle =s_\mu
\frac{1}{2\sqrt{3}} a^8, $$
and
$$
\langle p,s|A^0_\mu|p,s \rangle =s_\mu  a^0 ,
$$
where $p_\mu$ and $s_\mu$ are the momentum and the polarization
vector of the proton. $a^3$ and $a^8$ can be chosen $Q^2$
independent and may be determined from the $\frac{G_A}{G_B}$ and
$\frac{F}{D}$ ratios. $a^0(Q^2)$ evolves due to the anomaly and its
evolution can be described in the AB scheme \ct{spin} by
\bq
a^0(Q^2) = \Delta \Sigma - N_F \frac{\alpha_S(Q^2)}{2\pi} \Delta g(Q^2) .
\eq
Naive models or the $OZI$ approximation to $QCD$ lead at low energies to
\bq
a^0 \approx a^8 \approx 0.69 \pm 0.06 .
\eq
Experimentally\cite{forte}
\bq
a^0(\infty) = 0.10 \;^{+0.17}_{-0.10} .
\eq
The explanation for this unexpected small value has given rise to many
interpretations related to hadron structure, vacuum structure and evolution
\cite{ellis,kochelev,shore,altarelli}.

\subsection{The formalism}

To obtain the FSAC, we need to calculate the matrix elements of the
flavor singlet axial current. Let us write the current in the CBM
as a sum of two terms, one from the interior of the bag and the
other from the outside populated by the meson field $\eta^\prime$
(we will ignore the Goldstone pion fields for the moment; they will
be taken into account for the baryon charge leakage)
\bq
A^\mu =A^\mu_B \Theta_B + A^\mu_M \Theta_M.\label{current}
\eq
Since we will be dealing only with the flavor-singlet axial current, 
we will omit the flavor index in the current.
We shall use the short-hand notations $\Theta_B=\theta (R-r)$ and
$\Theta_M=\theta (r-R)$ with $R$ being the radius of the bag.
We demand that the $U_A (1)$ anomaly be given in this model by
\bq
\partial_\mu A^\mu =
\frac{\alpha_s N_F}{2\pi}\sum_a \vec{E}^a \cdot \vec{B}^a \Theta_{B}+
f m_\eta^2 \eta \Theta_{M}.\label{abj}
\eq
Our task is to construct the FSAC in the chiral bag model that is
gauge-invariant and consistent with this anomaly equation. Our
basic assumption is that in the nonperturbative sector outside of
the bag, the only relevant $U_A (1)$ degree of freedom is the
massive $\eta^\prime$ field. This assumption allows us to write
\bq
A^\mu_M = A^\mu_\eta = f\d^\mu \eta
\eq
with the divergence
\bq
\d_\mu A^\mu_\eta = fm_\eta^2 \eta.
\eq
Now the question is: what is the gauge-invariant and regularized
$A^\mu_B$ such that the anomaly (\ref{abj}) is satisfied? To
address this question, we rewrite the current (\ref{current}) as
\bq
A^\mu=A_{B_{Q}}^\mu + A_{B_{G}}^\mu + A_\eta^\mu\label{sep}
\eq
such that
\ba
\partial_\mu (A_{B_{Q}}^\mu + A_\eta^\mu) &=& f m_\eta^2 \eta
\Theta_{M},\label{Dbag}\\
\partial_\mu A_{B_G}^\mu &=&
\frac{\alpha_s N_F}{2\pi}\sum_a \vec{E}^a \cdot \vec{B}^a
\Theta_{B}. \label{Dmeson} \ea The subindices Q and G imply that
these currents are written in terms of quark and gluon fields
respectively. In writing (\ref{Dbag}), we have ignored the up and
down quark masses. We should stress that since we are dealing with
an interacting theory, there is no unique way to separate the
different contributions from the gluon, quark and $\eta$
components. In particular, the separation we adopt, (\ref{Dbag})
and (\ref{Dmeson}), is non-unique although the sum is without
ambiguity . We found however that this separation leads to a
natural partition of the contributions in the framework of the bag
description for the confinement mechanism that we are using here.

\subsubsection{The {\it quark} current $A_{B_Q}^\mu$}

The quark current is given by
\bq
 A^\mu_{B_Q} = \bar{\Psi} \gamma^\mu \gamma_5 \Psi
\eq where $\Psi$ should be understood to be the {\it bagged} quark
field. Therefore the quark current contribution to the FSAC is
given by \bq a^0_{B_Q} = \langle p|\int_B d^3r \bar{\Psi} \gamma_3
\gamma_5 \Psi|p\rangle. \la{aq}\eq The calculation of this type of
matrix elements in the CBM is nontrivial due to the baryon charge
leakage between the interior and the exterior through the Dirac
sea. But we know how to do this in an unambiguous way. A complete
account of such calculations can be found in \ct{pv,p,pr}. The
leakage produces an $R$ dependence which would otherwise not be
there in the matrix element of Eq.(\ref{aq}), as shown in Fig.~1.
It is significant that as seen in the figure there is no
contribution for zero radius, that is in the pure skyrmion
scenario for the proton. The contribution grows as a function of
$R$ towards the pure MIT result that would technically be reached
for infinite radius. The result of this calculation was first
presented in refs. \ct{pvrb,pv}. No new ingredient has been added.

\begin{figure}
%\vskip 2ex
\centerline{\epsfig{file=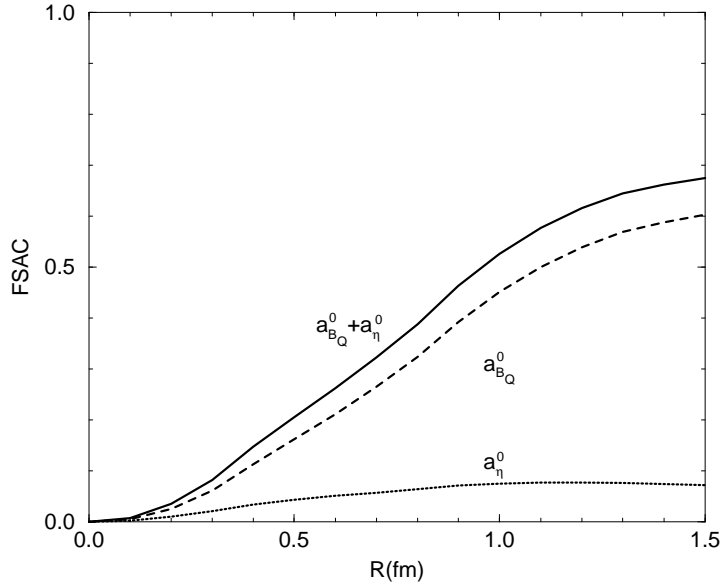, width=11cm}}
\caption{Various contributions to the flavor singlet axial current
of the proton as a function of bag radius : (a) quark  contribution
$A_{B_Q}$; (b) $\eta^\prime$ contribution $A_\eta$ and (c) the
sum.}
\end{figure}

\subsubsection{The {\it meson} current $A^\mu_\eta$}

Since we shall not add anything new to our previous result
obtained in \ct{pvrb,pv}, we will just quote the result. Due to
the coupling of the quark and $\eta$ fields at the surface, we can
simply write the $\eta$ contribution in terms of the quark
contribution, \bq a^0_\eta = \frac{1 +y_\eta}{2(1+y_\eta)
+y_\eta^2} \langle p|\int_B d^3r \bar{\Psi} \gamma_3 \gamma_5
\Psi|p\rangle. \la{aeta}\eq where $y_\eta = m_\eta R$. In Fig.~1
we show the radial dependence of this contribution, which arises
from the charge leakage mechanism, and follows the quark
distribution. Since the $\eta$ field has no topological structure,
its contribution also vanishes in the skyrmion limit.  This
illustrates how the dynamics of the exterior can be mapped to that
of the interior by boundary conditions. We may summarize the
analysis of these two contributions by stating that no trace of
the CCP is apparent in Fig.~1. Thus if the CCP were to emerge, the
only possibility would be that the gluons do the miracle!

\subsubsection{The {\it gluon } current $A^\mu_{B_G}$}

Understanding the FSAC and its implications in the present
framework involves crucially the role of the gluon contribution, in
particular its static properties and vacuum fluctuations, i.e., the
Casimir effects. The calculation of the Casimir effects constitutes
the principal aim of this work.

We begin by dividing the gluon current into two pieces
\bq
A^\mu_{B_G} = A^\mu_{G,stat}  + A^\mu_{G,vac}.
\eq
The first term arises from the quark and $\eta$ sources, while the
latter is associated with the properties of the vacuum of the
model. One might worry that this contribution could not be split in
these two terms without double counting. That there is no cause for
worry can be seen in several different ways. Technically, it is
easy to check it by noticing that the former acts on the quark Fock
space and the latter on the gluon vacuum. Thus, one can interprete
the former as a one gluon exchane correction to the quanity.
One can also show this intuitively by making the analogy
to the condensate expansion in QCD \ct{svz}, where the perturbative
terms and the vacuum condensates enter additively to the lowest
order.

Let us first describe the static term. We assume initially for
simplicity that there is no $\eta$ coupling. Then the boundary
conditions for the gluon field would correspond to the original
MIT ones \ct{mit}. The quark current is the  source term that
remains in the equations of motion after performing a perturbative
expansion in the QCD coupling constant, i.e., the quark color
current \bq g{\bar \Psi}_0 \gamma_\mu \lambda^a \Psi_0 \eq where
the $\Psi_0$ fields represent the lowest cavity modes. In this
lowest mode approximation, the color electric and magnetic fields
are given by \bq \vec{E}^a = g_s \frac{\lambda^a}{4\pi}
\frac{\hat{r}}{r^2} \rho (r) \label{ef} \eq \bq \vec{B}^a = g_s
\frac{\lambda^a}{4\pi}\left( \frac{\mu (r)}{r^3}(3 \hat{r}
\vec{\sigma} \cdot \hat{r} - \vec{\sigma}) + (\frac{\mu (R)}{R^3}
+ 2 M(r)) \vec{\sigma}\right) \label{bf} \eq where $\rho$ is
related to the quark density $\rho^\prime$ as\footnote{Note that
the quark density that figures here is associated with the color
charge, {\it not} with the quark number (or rather the baryon
charge) that leaks due to the hedgehog pion.}
\bq \rho
(r,\Gamma)=\int_\Gamma^r ds \rho^\prime
(s)\label{density}\nonumber \eq and $\mu, M$ to the vector current
density \ba \mu (r) &=& \int_0^r ds \mu^\prime (s),\nonumber\\ M
(r)&=& \int_r^R ds \frac{\mu^\prime (s)}{s^3}.\nonumber \ea The
lower limit $\Gamma$ is taken to be zero in the MIT bag model --
in which case the boundary condition is satisfied only {\it
globally}, that is, after averaging -- and $\Gamma=R$ in the so
called {\it monopole solution} \ct{pv,rv} -- in which case, the
boundary condition is satisfied {\it locally}.

After having clarified the procedure by recalling our old
calculation and pointing out the main difference of the present
with respect to that one, we proceed to introduce the $\eta$
field. We perform the same calculation with however the color
boundary conditions Eqs.(\ref{E}) and (\ref{B}) taken into
account. In the approximation of keeping the lowest non-trivial
term, the boundary conditions become \bq \hat{r}\cdot
\vec{E}_{stat}^a=-\frac{N_F g^2}{8\pi^2 f} \hat{r}\cdot
\vec{B}^a_g \eta (R)\label{Eg} \eq \bq \hat{r}\times
\vec{B}_{stat}^a=\frac{N_F g^2}{8\pi^2 f} \hat{r}\times
\vec{E}^a_g \eta(R).\label{Bg} \eq Here $\vec{E}^a_g$ and
$\vec{B}^a_g$ are the lowest order fields \ct{pv,rv} given by
(\ref{ef}) and (\ref{bf}) and $\eta (R)$ is the meson field at the
boundary. The $\eta$ field is given by
\bq \eta (\vec{r}) =
-\frac{g_{NN\eta}}{4\pi M} \vec{S} \cdot \hat{r}
 \frac{1+m_\eta r}{r^2} e^{-m_\eta r}
\la{eta}\eq
where the coupling constant is determined from the surface conditions
\ct{pv,rv}.

Note that the magnetic field is not affected by the new boundary
conditions, since $\vec{E}^a_g$ points into the radial direction.
The effect on the electric field is just a change in the charge,
i.e.,
\bq \rho_{stat}(r) = \rho(r,\Gamma) + \rho_\eta(R) \eq where
\bq \rho_\eta (R) = \frac{N_F g^2}{64 \pi^3 M}
\frac{g_{NN\eta}}{f} (1+y_\eta) e^{-y_\eta}. \eq

The contribution to the FSAC arising from these fields is
determined from the expectation value of the anomaly \bq
a^0_{G,stat} = \langle p|-\frac{N_F\alpha_s}{\pi} \int_B d^3r x_3
\vec{E}^a_{stat} \cdot \vec{B}^a_{stat}|p\rangle .
\label{astat}\eq The result of this contribution is shown in
Fig.~2, where we show the MIT solution, the { \it monopole} one
and the correction associated to both due to the color
coupling\footnote{We have also investigated electric fields of the
form $(\frac{A}{r^2} + Br)\hat{r}$, but the results do not change
much with respect to the ones shown since the $B$ term tends to be
small.}. One sees that including the $\eta$ contribution in
$\rho_{stat} (r)$ brings a non-negligible modification to the FSAC
but does not modify the result qualitatively. What is most
striking is the drastic difference between the effect of the
MIT-like electric field and that of the monopole-like electric
field: The former is totally incompatible with the Cheshire Cat
property whereas the latter remains consistent independently of
whether or not the $\eta$ contribution is included in
$\rho_{stat}$.

\begin{figure}
%\vskip 2ex
\centerline{\epsfig{file=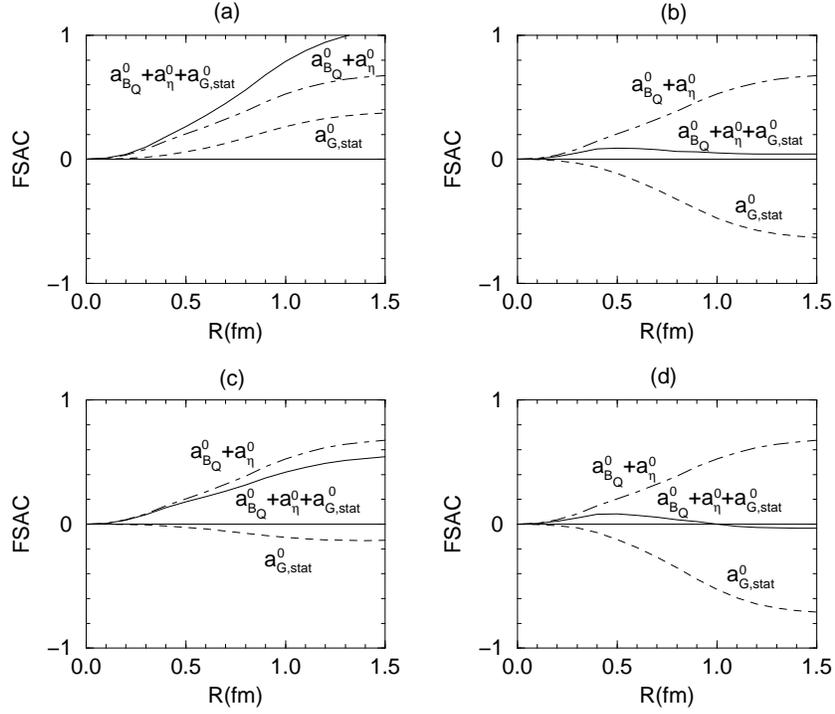, width=11cm}}
\caption{Dependence of $a_{G,stat}^0$ on the choice of $\Gamma$
and the boundary conditions as a function of bag radius : (a) with
an MIT-like electric field without $\eta$ coupling, (b) with a
{\it monopole}-like electric field without $\eta$ coupling, (c)
with an MIT-like electric field {\em with} $\eta$ coupling, and
(d) with a {\it monopole}-like electric field {\em with} $\eta$
coupling.}

\end{figure}

Finally the $A_{G,vac}$ term arises from the so called Casimir
effect of the anomaly term. The vacuum in the cavity and the
perturbative vacuum are different due to the geometry of the
cavity. This effect has been considered for many other observables
and also for the quarks in this calculation
\ct{p,pr,pv}, but never for the gluons. We proceed in the next
section to describe this calculation.

\section{The Gluon Casimir Calculation}

The quantity that we wish to calculate is the gluon vacuum
contribution to the flavor singlet axial current of the proton. It
can be done by evaluating the expectation value
\begin{equation}
\langle 0_B| -\frac{N_F\alpha_s}{\pi} \int_V d^3r
x_3 ( \vec{E}^a \cdot \vec{ B}^a) |0_B\rangle
\label{f1}\end{equation}
where $|0_B\rangle$ denotes the vacuum in the bag. The standard way
to evaluate this expectation value would be to expand the field
operators in terms of the classical eigenmodes that satisfy the
equations of motion and the boundary conditions. Although
well-defined, this approach is technically involved. We have not
yet obtained any quantitative results to report. In this paper, we
shall proceed in the opposite direction. Instead of arriving at the
CCP as in the standard approach, we shall {\it assume} the CCP and
evaluate the Casimir contribution with the expression that follows
from the assumption. The idea goes as follows.

The CCP states that at low energy, hadronic phenomena do not
discriminate between QCD degrees of freedom (quarks and gluons) on
the one hand and meson degrees of freedom (pions, etas,...) on the
other, provided that all necessary quantum effects (e.g., quantum
anomalies) are properly taken into account. If we consider the
limit where the $\eta$ excitation is a long wavelength oscillation
of zero frequency, the CCP asserts that it does not matter whether
we choose to describe the $\eta$, in the interior of the
infinitesimal bag, in terms of quarks and gluons or in terms of
mesonic degrees of freedom. This statement, together with the
color boundary conditions, leads to an extremely simple and useful {\it
local} formula \ct{nrwz1},
 \begin{equation}
\vec{ E}^a \cdot \vec{ B}^a \approx -\frac{N_F g^2}{8\pi^2}
\frac{\eta}{f} \frac12 G^2, \label{f2}\end{equation} where only
the term up to the first order in $\eta$ is retained in the
right-hand side. Here we adapt this formula to the CBM. This means
that the couplings are to be understood as the average bag
couplings and the gluon fields are to be expressed in the cavity
vacuum through a mode expansion. In fact, by comparing the
expression for the $\eta^\prime$ mass derived in \ct{nrwz1} using
Eq.(\ref{f2}) with that obtained by Novikov et al \ct{NSVZ} in QCD
sum-rule method, we note that the matrix element of the $G^2$ in
(\ref{f2}) should be evaluated in the {\it absence} of light
quarks. This means, in the bag model, the cavity vacuum. That the
surface boundary condition can be interpreted as a local operator
is a rather strong CCP assumption which while justifiable for
small bag radius, can only be validated a posteriori by the
consistency of the result. This procedure is the substitute to the
condensates in the conventional discussion.

Substituting Eq.(\ref{f2}) into Eq.(\ref{f1}) we obtain
\begin{eqnarray}&& \hskip -2em
\langle 0_B| -\frac{N_F\alpha_s}{\pi} \int_V d^3r x_3 (\vec{ E}^a
\cdot \vec{ B}^a) |0_B\rangle \nonumber\\ &&
\approx
\left(-\frac{N_F\alpha_s}{\pi} \right)
\left(-\frac{N_F g^2}{8\pi^2} \right) \frac{y(R)}{f_0}
\langle p|S_3|p\rangle
\langle 0_B |\int_V d^3r \frac12 G^2 x_3 \hat{x}_3 |0_B\rangle \nonumber \\
&& \approx  \left(-\frac{N_F\alpha_s}{\pi} \right)
\left(-\frac{N_F g^2}{8\pi^2} \right) \frac{y(R)}{f_0}
\langle p|S_3|p\rangle (N_c^2-1) \nonumber\\
&& \hskip 3em
\sum_n \int_V d^3r (\vec{ B}_n^* \cdot \vec{ B}_n -
\vec{ E}_n^* \cdot \vec{ E}_n) x_3 \hat{x}_3 ,
\label{ps1}\end{eqnarray}
where we have used that $\eta$ has a structure of $(\vec{
S}\cdot\hat{ r}) y(R)$. Since we are interested only in the first
order perturbation, the field operator can be expanded by using MIT
bag eigenmodes (the zeroth order solution). Thus, the summation
runs over all the classical MIT bag eigenmodes. The factor
$(N_c^2-1)$ comes from the sum over the abelianized gluons.

The next steps are the  numerical calculations to evaluate the mode
sum appearing in Eq.(\ref{ps1}): (i) introduction of the heat
kernel regularization factor to classify the divergences appearing
in the sum and (ii) subtraction of the ultraviolet divergences.
\vskip 1ex

\subsection {Normalization of the eigenmodes}

The classical eigenmode of the (abelianized) gluons confined in
the MIT bag can be classified by the total spin quantum numbers
$(J,M)$ given by the vector sum of the orbital angular momentum
$\vec{ L}$ and the spin $\vec{ S}$,
\begin{equation}
\vec{ J}\equiv \vec{ L} + \vec{ S},
\end{equation}
and the radial quantum number $n$. There are two kinds of classical
eigenmodes according to the relations between the parity and the
total spin; (i) M-mode with the parity $\pi=-(-1)^{J}$ and (ii)
E-mode with the parity $\pi = -(-1)^{J+1}$. Here, the extra minus
sign is due to the negative intrinsic parity of gluon.

We will work with the vector fields with the gauge choice,
\begin{equation}
G_0 = 0, \mbox{ and } \vec{\nabla}\cdot \vec{ G} = 0.
\end{equation}
Then the electric field and the magnetic field are obtained through
the relations
\begin{eqnarray}
\vec{ E} &=& -\frac{\partial \vec{ G}}{\partial t}, \\
\vec{ B} &=& \vec{\nabla} \times \vec{ G}.
\end{eqnarray}

Explicitly, the solutions are obtained as
\begin{eqnarray}
\mbox{(i) M-modes : }&& \nonumber\\
&&\vec{ G}^{(M)}_{(n,J,M)}(\vec{ r}) = {\cal N}_M j_{J}(\omega_n r)
\vec{ Y}_{J,J,M}(\hat{ r}),
\label{A_M} \\
 \mbox{(ii) E-modes : }&& \nonumber\\
&&\vec{ G}^{(E)}_{(n,J,M)}(\vec{ r}) = {\cal N}_E \left[
-\sqrt{\frac{J}{2J+1}} j_{J+1}(\omega_n r) \vec{ Y}_{J,J+1,M}(\hat{ r})
\right. \\
&& \hskip 9em \left.
+\sqrt{\frac{J+1}{2J+1}} j_{J-1}(\omega_n r) \vec{ Y}_{J,J-1,M}(\hat{ r})
\right],
\label{A_E}
\end{eqnarray}
where $\vec{ Y}_{J,\ell,M}$ is the vector spherical harmonics of
the total spin $J$ composed of the angular momentum $\ell$ and
$j_{\ell}(x)$ is the spherical Bessel functions. The energy
eigenvalues are determined to satisfy the MIT boundary conditions
as
\begin{eqnarray}
 \mbox{(i) M-modes : }&&\nonumber\\
&& X_n j^\prime_J(X_n) + j_J(X_n) = 0, \\
\mbox{(ii) E-modes : }&&\nonumber\\
&& j_J(X_n) = 0
\end{eqnarray}
where we have defined $X_n=\omega_n R$.
The normalization constants ${\cal N}_{M,E}$ will be specified
below.

The field operator $\vec{ G}(\vec{ r},t)$ is expanded in terms of
the classical eigenmodes as
\begin{equation}
\vec{ G}(\vec{ r},t) = \sum_{ \{\nu\} }\left( a_{ \{\nu\} }
\vec{G}_{ \{\nu\} }(\vec{ r}) e^{-i\omega_n t} +a^\dagger_{
\{\nu\} } \vec{ G}^*_{ \{\nu\} }(\vec{ r}) e^{+i\omega_n t}
\right), \label{A}\end{equation} where $\{\nu\}$ denotes the
quantum number set $(n,J,M,\lambda=$E or M).

We determine the normalization constants ${\cal N}_{M,E}$ in such a way that
the free gluon Hamiltonian operator
\begin{equation}
H =  \frac12\int_B d^3 r  (\vec{E}\cdot \vec{ E} + \vec{ B} \cdot \vec{ B} )
\label{H}\end{equation}
becomes
\begin{equation}
H = \sum_{\{\nu\}} \omega_{\{\nu\}} a^{\dagger}_{\{\nu\}}  a_{\{\nu\}}
\end{equation}
when Eq.(\ref{A}) is substituted into Eq. ({\ref{H}).
It leads to a normalization condition for the classical eigenmodes as
\begin{equation}
\int_B d^3r \vec{ G}^*_{\{\nu\}} \cdot \vec{ G}_{\{\mu\}}
 = \frac{1}{2\omega_{\{\nu\}}} \delta_{\{\nu\}\{\mu\}}.
\end{equation}
Then the normalization constants are determined explicitly as
\begin{eqnarray}
{\cal N}_M &=& \left[ X_n R^2 \left( j^2_J(X_n) - j_{J-1}(X_n)
j_{J+1}(X_n) \right) \right]^{-1/2}, \\
{\cal N}_E &=& \left[ X_n R^2 j_{J-1}^2(X_n) \right]^{-1/2}.
\end{eqnarray}
\vskip 1ex

\subsection{ Matrix elements}
The first step is to calculate the matrix elements
\begin{equation}
Q_{\{\nu\}} \equiv  \int_B d^3 r
( \vec{ B}^*_{\{\nu\}} \cdot \vec{ B}_{\{\nu\}}
- \vec{ E}^*_{\{\nu\}} \cdot \vec{ E}_{\{\nu\}} ) x_3 \hat{x}_3.
\end{equation}
>From Eq.(\ref{A_M}), we obtain
\begin{eqnarray}
\vec{ E}_{\{\nu\}}(\vec{ r}) & = & (+i\omega_n) {\cal N}_M j_J(\omega_n r)
\vec{ Y}_{J,J,M}(\hat{ r}), \\
\vec{ B}_{\{\nu\}}(\vec{ r}) & = & (+i\omega_n) {\cal N}_M \left[
-\sqrt{\frac{J}{2J+1}} j_{J+1}(\omega_n r) \vec{ Y}_{J,J+1,M}(\hat{ r})
\right. \\
&& \hskip 9em \left.
+\sqrt{\frac{J+1}{2J+1}} j_{J-1}(\omega_n r) \vec{ Y}_{J,J-1,M}(\hat{ r})
\right],
\end{eqnarray}
for the M-modes and the similar equations with $\vec{ E}$ and $\vec{ B}$
being interchanged for the E-modes.

We encounter in the calculation the following angular integrals
\begin{equation}
\int d\Omega \vec{ Y}_{J,\ell,M}^* \cdot \vec{ Y}_{J,\ell,M} \hat{x}_3^2.
\end{equation}
By using that $\hat{x}_3^2 = (4/3)\sqrt{\pi/5} Y_{20} + 1/3$ and
the Wigner-Eckart theorem, we obtain
\begin{equation}
\int d\Omega \vec{ Y}_{J,\ell,M}^* \cdot \vec{ Y}_{J,\ell,M} \hat{x}_3^2
= c_{J,\ell} (J(J+1)-3M^2) + \frac{1}{3},
\end{equation}
where $c_{J,\ell}$ is a constant that depends only on $J$ and $\ell$.
We have to perform  the summation over $M$, which runs from $-J$ to $J$,
which cancels the contribution of the first term, therefore we can take
{\it effectively} 1/3 as the result of the integral.

Finally, we obtain the matrix elements for the M-modes as
\begin{equation}
Q^{(M)}_n = \frac{1}{3} \frac{\int^{X_n}_0 x^3 dx \left[ j_J^2(x)
- \frac{J}{2J+1} j_{J+1}^2(x) - \frac{J+1}{2J+1} j_{J-1}^2(x)
\right] } { X_n^3 \left[ j^2_J(X_n) - j_{J-1}(X_n) j_{J+1}(X_n)
\right] }.
\end{equation}
In the case of the E-mode, we obtain exactly the same formula
except the minus sign in front of it. (Note that the formulas for
the electric field and the magnetic field are interchanged.)

\begin{figure}
%\vskip 2ex
\centerline{\epsfig{file=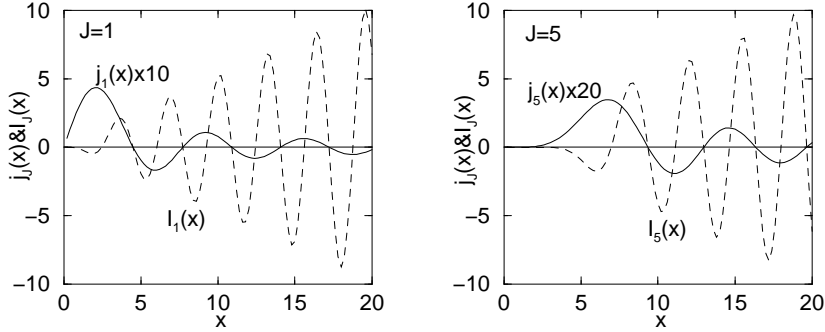, width=11cm}}
\caption{$j_J(x)$ and $I(x)$ as a function of x.}
\end{figure}

We have found that the matrix elements for the E-mode vanish
up to our numerical accuracy as shown in Fig.~3. Here, the solid
line is the spherical Bessel function $j_J(x)$ and the dashed line
is the integral
\begin{equation}
I(x) \equiv \int^{x}_0 y^3 dy \left[ j_J^2(y) -
\frac{J}{2J+1} j_{J+1}^2(y) - \frac{J+1}{2J+1} j_{J-1}^2(y) \right]
\end{equation}
We see that the zeroes of $I(x)$ and $j(x)$ coincide, thus showing
that $Q_n^{(E)} (X_n)=0$. We have been unable however to prove
this result analytically, except the trivial case of $J=0$. \vskip 1ex

\subsection{The mode sum}

\begin{figure}
%\vskip 2ex
\centerline{\epsfig{file=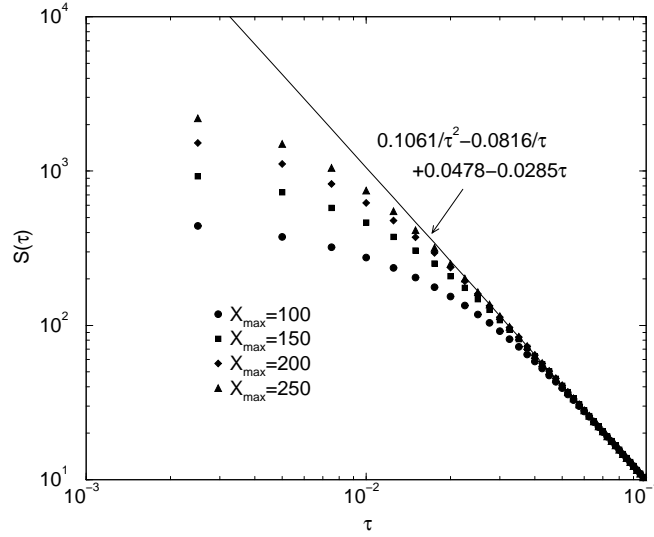, width=10cm}}
\caption{Diverging properties of $S(\tau)$ as a function of the
heat kernel regularization parameter $\tau$. All the magnetic modes
up to $\omega_n R(\equiv X_n)$=100(solid circle), 150(solid square),
200(solid diamond) and 250(solid triangle) are included in the sum.}
\end{figure}

 In order to regularize the mode sum,
we introduce a heat kernel factor $\exp(-\tau X_n)$;
\begin{equation}
S(\tau) \equiv \sum_{n,J} (2J+1) Q^{(M)}_{n,J} e^{-\tau X_n},
\end{equation}
where we have carried out the trivial sum over $M$ and the vanishing E-mode
contribution is excluded.

Fig.~4 shows the numerical results of the sum up to $X_{max}$=100,
150, 200, 250 for the 40 values of $\tau$ from 0.0025 to 0.1 with
the step 0.0025. We can see that below $\tau < 0.06$ the
convergence is poor. However, it is enough to see the presence of
an $1/\tau^2$ divergence. If we fit the data above $\tau
> 0.06$, we obtain
\begin{equation}
S(\tau) = \frac{0.1061}{\tau^2} - \frac{0.0816}{\tau} + 0.0478 -
0.0285\tau. \label{nf}\end{equation} Apart from a possible
logarithmic divergence, there are quadratic and linear divergences
as we set $\tau$ equal to zero . We shall remove these divergences
following a procedure commonly used in Casimir problems
\cite{schwinger}. Caveat on this procedure will be highlighted in
the discussion section. Now if we neglect logarithmic divergences
that might be present , the best way to get rid of the quadratic
and linear divergences is to evaluate
\begin{equation}
S(\tau)+2\tau S^\prime(\tau)+\frac12\tau^2 S^{\prime\prime}(\tau)
= \sum_{n,J} (2J+1) Q_{n,J} (1-2\tau X_n+0.5\tau^2 X_n^2) e^{-\tau X_n}.
\end{equation}

Fig.~5 show the results on this quantity for 80 values of $\tau$
ranging from 0.0025 to 1.
We see that no serious divergences appear
anymore. By fitting the convergent data with the above expressions for
 $\tau$,
we obtain for the finite part of the sum  0.0478, from the cubic
function fit, and 0.0456, from the quadratic one. These results
are comparable to the finite term of the above naive fitting
procedure (\ref{nf}), which yielded 0.0478.
\begin{figure}
%\vskip 2ex
\centerline{\epsfig{file=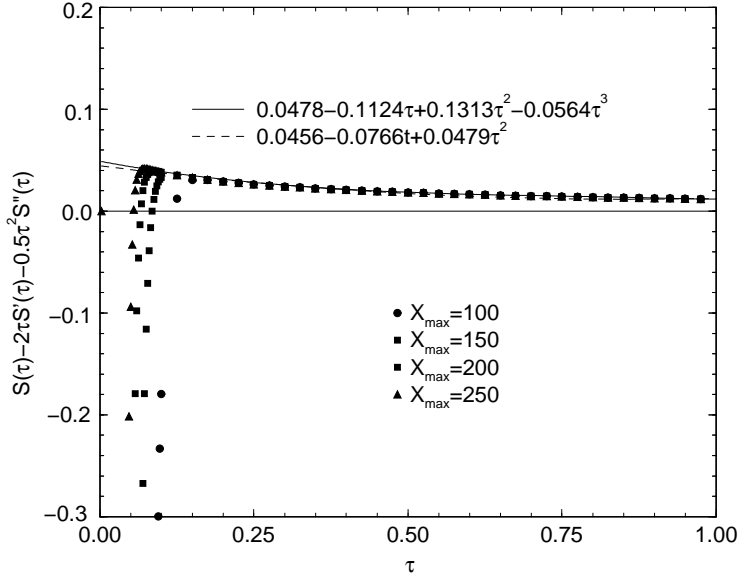, width=11cm}}
\caption{$S(\tau)-2\tau S^\prime(\tau)+\frac12\tau^2 S^{\prime\prime}(\tau)$
as a function of $\tau$. The finite term of $S(\tau)$ is  extracted by
fitting these quantities to a cubic and quadratic curves.}
\end{figure}

Once we have the numerical value on the mode sum, the gluon vacuum
contribution to FSAC can be evaluated simply as
\begin{equation}
a_{G,vac}^0 = -\frac{(2.10)^2}{2}\times\frac{8}{2} \times
\frac{y(R)}{122\mbox{MeV}} \times (0.0478),
\end{equation}
where $y(R)$ is related to $a^0_{B_Q}$ as
\begin{equation}
 y(R) = -\frac{3m_\eta^2}{8\pi f_\eta}
\frac{(1+m_\eta R)}{[2(1+m_\eta R) + (m_\eta R)^2](m_\eta R)^2}
a^0_{B_Q}.
\end{equation}
We have used $N_F=N_c=3$, $\alpha_s=2.2$,
$f_0=\sqrt{N_F/2}f_{\eta^\prime} \sim 122$MeV and $m_\eta=958$ MeV.

\begin{figure}
%\vskip 2ex
\centerline{\epsfig{file=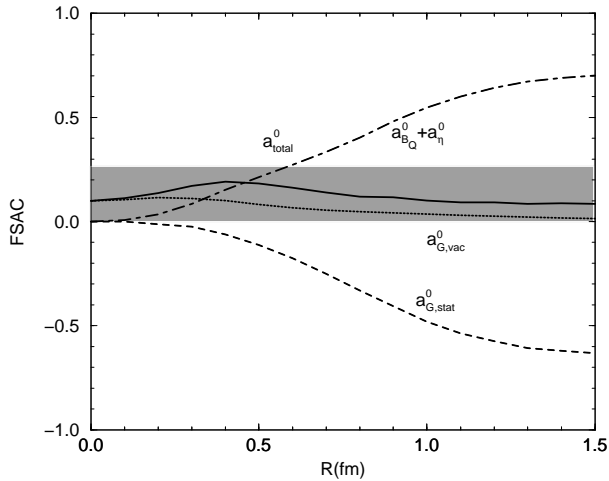, width=9cm}}
\caption{Various contributions to the flavor singlet axial current
of the proton as a function of bag radius and comparison with the
experiment: (a) quark plus $\eta$ contribution ($a^0_{B_Q} +
a^0_\eta$), (b) the contribution of the static gluons due to quark
source ($a^0_{G,stat}$), (c) the gluon vacuum contribution
($a^0_{G,vac}$), and (d) their sum ($a^0_{total})$. The shaded area
corresponds to the range admitted by experiments.}\label{result}
\end{figure}
\section{Results and Conclusions}
Our numerical results are given in Fig.\ref{result}. Standard MIT bag
parameters were used for the calculation. The quarkish component of
the FSAC is given by the sum of the quark and $\eta$ contributions,
$a^0_{B_Q}+a^0_\eta$ and the gluonic component by $a^0_{G,stat}+a^0_{G,vac}$.
Both increase individually as the confinement size $R$ is increased
but the sum remains small, $0<a^0_{total}<0.3$ for the whole range of
radii, consistent with the experiment,
$a^{exp}=a^0(\infty)=0.10^{+0.17}_{-0.10}$. It is remarkable that
$a(R=0)\simeq a(R\approx 1.5\ {\rm fm})$ while each component can
differ widely for the two extreme radii.

We have shown that the principal agent for the observed small FSAC
in the proton in the framework of the chiral bag model is the
Cheshire Cat phenomenon (CCP). It is the CCP that assures the
cancellation between two contributions, one from the quarkish
component and the other from the gluonic component in the
particular way the separation is made. For a small bag radius, both
components are small, so the net FSAC being small is immediate.
This is consistent with the observation that in the limit that
$R\rightarrow 0$, we recover the skyrmion description which gives a
vanishing FSAC at the leading order, modified by matter fields at
the next order. At large bag radius which leads to the MIT bag
model, both the quarkish contribution and the gluonic contribution
are large but they cancel. Our assertion is that this cancellation
is caused by the CCP. We should however recall that the separation
between the quarkish component and the gluonic component we adopted
in (\ref{Dbag}) and (\ref{Dmeson}) is entirely arbitrary although
the sum is unique. {\it Whether the separate component by itself is
large or small has no physical meaning. Only the total does.}
Different separations would lead to different scenarios leading to
the same small value. It is plausible that in some limit -- unknown
to us -- the FSAC would be exactly zero with the finite nonzero
value indicating a departure from this limit. Understanding this
limit would allow a unique separation of the components.

One of the principal results of this paper is that {\it it is
possible to have
 a nonzero value for the FSAC
at $R=0$ and is of the same size as at large $R$\footnote{The
reason for this nonzero value is intimately connected with the
CCP, since it is the finite part of the gluon mode sum which
normalizes the value of this contribution at the origin. Moreover
the color boundary condition provides us with a decreasing $\eta$
field contribution which changes softly as a function of $R$.}}.
While the effect of the surface color anomaly term is generally
small for all radii, the finite nonzero value of FSAC for $R=0$ is
assured by the surface boundary term. Thus the violation of the
CCP observed in the previous calculations at $R=0$ \cite{pv,rv} is
neatly eliminated by the color anomaly boundary condition. More
importantly, the {\it monopole} structure of the color electric
field previously proposed is found to be required for the sign
that comes with the important static gluonic contribution from the
quark source. We believe that this cancellation is a manifestation
in the bag scenario of the recently discovered one for $QCD$
\cite{kochelev}. The MIT configuration would strongly violate the
CCP. We are thus led to the conclusion that the CCP {\it requires}
the monopole configuration for the color electric field. Whether
or not this configuration leaves undisturbed other -- successful
-- phenomenology was discussed in \cite{pv}.

In calculating the gluonic Casimir effect, we {\it made} the {\it
ab initio} assumption that the CCP holds, an assumption which is
expected to be valid for small bag radius. We then extend it, in
accordance with the CCP, to all bag radii. We can justify this only
a posteriori by showing that the CCP assumption is consistent with
what one gets out. Note however that the gluonic Casimir effect is
most significant for small $R$ where it is needed for the CCP and
plays little role for large $R$. Thus our assumption is validated.
It would of course be more satisfying if one could obtain the CCP
as an output of the formalism, not put in as an input. Such a
calculation is in progress.

We should mention a caveat left unspecified in the text in
regularizing this Casimir contribution. Since $a_{G,vac}^0$
vanishes when the $\eta$ field is removed, the so-called ``vacuum
contribution" is duly subtracted in what we have computed. However
we have also explicitly subtracted quadratic and linear
divergences appearing from the mode sum by resorting to a
procedure used in the past in most of Casimir-type calculations
\cite{schwinger} which as far as we know, is physically reasonable
but has not yet been rigorously justified from first principles.
The same caveat applies to our calculation as it does to others.
The finite term we have obtained might therefore be subject to
additional finite corrections. In this paper we have invoked the
Cheshire Cat Principle to ignore such corrections in
$a_{G,vac}^0$. We hope that the calculation in progress
\cite{lmprv} will eliminate this ambiguity.

Given the caveat mentioned above and the approximations used, our
result can at best be qualitative. A better treatment (such as a
more realistic gauge coupling constant running with the bag size,
a more accurate calculation of $a^0_{G,vac}$ etc.) might modify
the result quantitatively. Even so, we believe it to be quite
robust that the overall FSAC is small, $\lsim 0.3$ and that it is
more or less independent of the confinement size.

\section*{Acknowledgments}

This work started as a pilot activity in astro-hadron physics at
Korea Institute for Advanced Study (KIAS) when two of us (MR
\& VV) were visiting the Institute in February 1998.
They would like to thank KIAS for support and C.W. Kim, Director of
KIAS, for his efforts to make the collaboration a most profitable
and pleasant one. VV also benefited from a travel grant from the
Conselleria de Cultura, Educaci\'o i Ci\`encia de la Generalitat
Valenciana, which allowed his visit to DAPNIA-CEA Saclay, where
part of this work was done. He thanks J.P. Guichon and J.M. Laget
for their efforts to make the visit possible and the members of the
Service for the hospitality. His work was partially supported by
DGICYT-PB97-1227 and by the TMR programme of the European
Commission ERB FMRX-CT96-008.
The work of HJL, DPM and BYP is partially supported by the Korea Science
and Engineering Foundation through CTP of Seoul National University.

\end{document}